\documentclass{iaus}
\usepackage{graphicx}
\usepackage{natbib}

\title[Shocks, CMEs \& Flare effects on Space Weather] 
{Coronal Shocks Associated with CME and Flares and their Space Weather Consequences \\ }

\author[Marina Laskari \& al]   
{Marina Laskari$^1$,
Panagiota Preka-Papadema$^1$,
Constantine Caroubalos$^2$,
George Pothitakis$^2$,
Xenophon Moussas$^1$,
Eleftheria Mitsakou$^1$
\and A. Hillaris$^1$}

\affiliation{
$^1$Department of Physics, University of Athens, 15784 Athens, Greece\\[\affilskip]
$^2$Department of Informatics, University of Athens, 15783 Athens, Greece}

\pubyear{2008}
\volume{257}  
\jname{Universal Heliophysical Processes}
\editors{A.C. Editor, B.D. Editor \& C.E. Editor, eds.}
\begin{document}

\maketitle

\begin{abstract}
We study the geoeffectiveness of a sample of complex events;
each includes a coronal type II burst, accompanied by a GOES SXR flare and LASCO CME. 
The radio bursts were recorded by the ARTEMIS-IV radio spectrograph, in the 100-650 MHz range;
the GOES SXR flares and SOHO/LASCO CMEs, were obtained from the Solar Geophysical Data (SGD)
and the LASCO catalogue respectively. These are compared with changes of solar
wind parameters and geomagnetic indices in order to establish a relationship between solar energetic events
and their effects on geomagnetic activity.

\end{abstract}

\begin{figure}[p]
\begin{center}
 \includegraphics{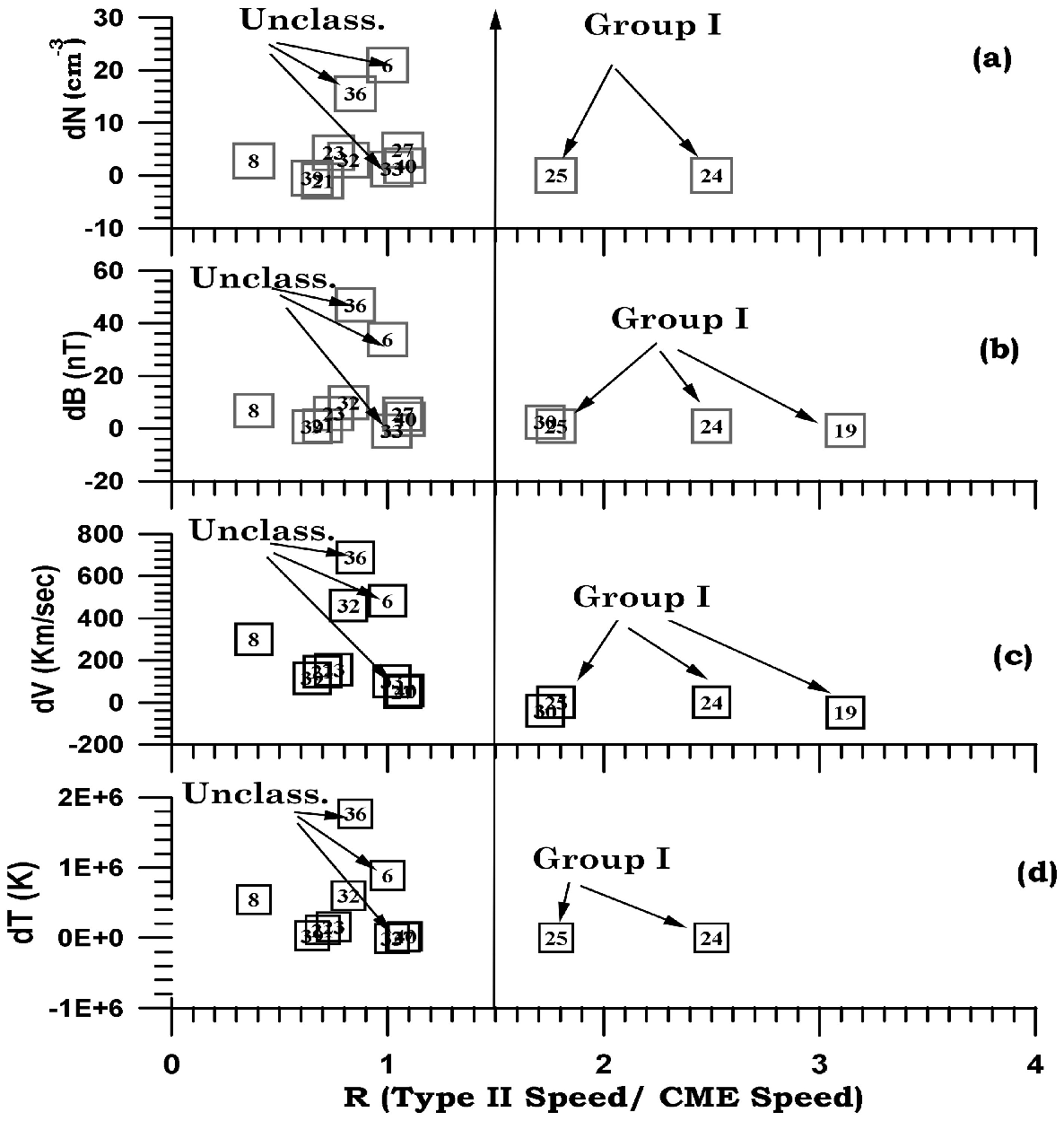}
 \includegraphics{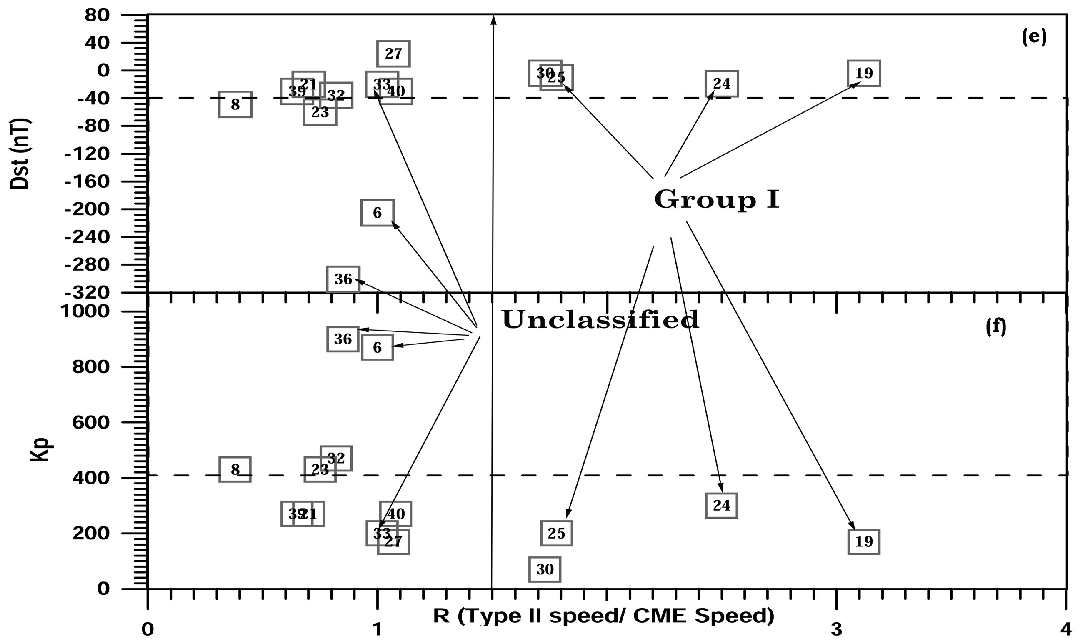}
 \caption{Solar Wind Parameters Variation \& Space Weather indices as a function of the ratio of the speed
 of Type II to the CME speed ($V_r$=$V_{II}/V_{CME}$): (a)Density (dN) in $cm^{-3}$, 
(b) Magnetic Field (dB) in nT, (c) Speed  (dV) in Km/sec, (d) Temperature (dT) in K, (e) Dst, (f) Kp.
 (We have labeled Events of Group I \& Unclassified (cf. Table 1 of Pothitakis et al in these Proceedings);
   The quiet Solar Wind Parameters at 1AU were: B=5 nT, V=350 Km/sec, N=5 $cm^{-3}$, T=$10^5$K.)}
   \label{fig2}
\end{center}
\end{figure}

\firstsection 

\section{Introduction}

The primary sources of geomagnetic phenomena are the solar eruptive events; they initiate the disturbances of
solar wind parameters (magnetic field, speed, density and temperature) which in turn drive the Space Weather
effects. The latter is characterized by a geomagnetic index such as the {\textit{Disturbance storm time}}
or {\bf{Dst}} (cf. \citet{Gopalswamy07}; the geomagnetic field variations are, on the other hand, quantified, 
on average, by the {\bf{Kp}} index. Though the connection of the geomagnetic phenomena with processes on
the Sun is well established the geomagnetic storm effectiveness of CMEs and solar flares  is still an open 
question as published results are not conclusive (for a review cf. \citet{Yermolaev05},
also \citet{Yermolaev06}).

We study the geoeffectiveness of a medium size sample of solar events; each includes a
coronal type II burst sometimes extending to an interplanetary type II, accompanied by a GOES SXR
flare and SOHO/LASCO CME. The radio bursts were recorded by the ARTEMIS-IV radio
spectrograph (\cite{Caroubalos01}); the GOES SXR flares and SOHO/LASCO CMEs, were obtained from the Solar
Geophysical Data (SGD) and the SOHO/LASCO lists respectively.

We examine the effects on the solar wind parameters and the {\bf{Kp}}, {\bf{Dst}} geomagnetic indices variations two
to three days after the events recording.

\section{Data selection \& Analysis}

The ARTEMIS IV radiospectrograph (\cite{Caroubalos01})
observed about 40 type II and/or IV radio bursts which they were published in the form of a
catalogue \citep{Caroubalos04}. From this catalogue we have used in our study the same fourteen events 
studied by Pothitakis et al (these proceedings); here we adopted the same numbering of events.
The solar wind parameters (magnetic field (B and Bz component), speed (V), temperature (T) and density (N))
at 1 A.U., and the geomagnetic indices Kp and Dst were obtained from the OMNI data base.

The data set was carefully selected in order to represent solar events within periods of relative inactivity
(deduced from the GOES SXR flux 3-4 days prior and after the event) in order to establish an association
between the space weather effects and the solar driver.

In Pothitakis et al (cf. their Table 1) the association of the flare--CME--Type II burst parameters was
studied; in this report on the other hand, we examine the variation of the geomagnetic indices {\bf{Dst}}
\& {\bf{Kp}} and of the Solar Wind Parameters (dB, dV, dT, dN)) with the ratio of the speed of Type II to
the CME speed ($V_r$=$V_{II}/V_{CME}$); this ratio provides a sort of indication of the CME capability
in driving a magnetohydrodynamic shock. The results of this comparison are plotted in figure \ref{fig1}.

Our examination indicates that a certain class of events may initiate space storms
(Events 23, 06, 36, 08 \& 32) give {\bf{Dst}}$<$-40 or {\bf{Kp}}$>$400), the Bastille Event (36) among 
them.  

Comparing this result with the classification of Pothitakis et al we note that the most geoeffective events
of our data sample are from Group II (23, 08, 32) plus two of the events which were not classified (36, 06).
The events in Group I (24, 25, 19), on the other hand, include fast coronal shocks (1213--1940 km/sec) and CMEs with
speeds in the 400--800 km/sec range yet their effect on space weather are small.

Lastly, we note that there is a trend of increased geoeffectiveness and Solar Wind Variations with $V_r$ but only when $V_r$ $<$ 1.5.

\section{Final Remarks}

We have studied the geoeffectiveness of a medium size sample of complex events, each including
a coronal type II burst, a GOES SXR flare and a SOHO/LASCO CME; the results were compared with
the classification of Pothitakis et al (these proceedings) which was based on
parameters, related to shock \& CME kinetics and radio bursts-flare-CME timing.

Fast coronal shocks do not, always, initiate storms, yet fast shocks in the front of fast CMEs
have a higher probability of inciting disturbances in the near earth environment. Further more,
a trend of increase in geomagnetic effects with R was found when $V_r$ $<$ 1.5. This, in fact, excludes
fast coronal shocks accompanied by relatively slow CMEs.

\end{document}